\documentclass[journal,10pt,twocolumn,comsoc]{IEEEtran}
\IEEEoverridecommandlockouts
\usepackage[colorlinks,citecolor=blue,linkcolor=blue,urlcolor=blue]{hyperref}
\usepackage{multicol}
\usepackage{cite}
\usepackage{multibib}
\newcites{Apndx}{References}
\usepackage{upgreek}
\usepackage{graphicx}
\usepackage{amsmath,amssymb,amsfonts}
\usepackage{bbold}
\usepackage{algorithmic}
\usepackage{gensymb}
\usepackage{color}
\usepackage{fancyhdr}
\newcommand{\ket}[1]{|#1\rangle}

\newcommand{\braket}[2]{\langle#1|#2\rangle}
\newcommand{\ensavg}[1]{\langle#1\rangle}

\newcommand{\DISTRIBUTION}{Approved for Public Release; Distribution is Unlimited; \#21-1020; Dated 06/24/21}

\begin{document}
\title{Temporal Modes of Light in Satellite-to-Earth \\ Quantum Communications}
\author{Ziqing Wang, Robert Malaney, and~Ryan~Aguinaldo
\thanks{Ziqing Wang and Robert Malaney are with the School of Electrical Engineering \& Telecommunications, The University of New South Wales, Sydney, NSW 2052, Australia (e-mail: \{ziqing.wang1, r.malaney\}@unsw.edu.au). Ryan Aguinaldo is with Northrop Grumman Mission Systems, San Diego, California, USA (e-mail: Ryan.Aguinaldo@ngc.com).}
\vspace{-0.4in}
\vspace{5pt}
}

\maketitle
\thispagestyle{fancy}

\begin{abstract}

The photonic Temporal Mode (TM) represents a possible candidate for the delivery of viable multidimensional quantum communications. However, relative to other multidimensional quantum information carriers such as the Orbital Angular Momentum (OAM), the TM has received less attention. Moreover, in the context of the emerging quantum internet and satellite-based quantum communications, the TM  has received no attention.
In this work, we remedy this situation by considering the traversal through the satellite-to-Earth channel of single photons encoded in TM space. Our results indicate that for anticipated atmospheric conditions the photonic TM offers a promising avenue for the delivery of high-throughput quantum communications from a satellite to a terrestrial receiver. In particular, we show how these modes can provide for improved multiplexing performance and superior quantum key distribution in the satellite-to-Earth channel
, relative to OAM single-photon states.
The levels of TM discrimination
that guarantee this outcome are outlined and implications of our results for the emerging satellite-based quantum internet are discussed.
\end{abstract}

\vspace{-0.55cm}
\section{Introduction}
The vision of global-scale and highly-secure quantum communication networks has been turned into reality with the help of satellite-to-Earth Quantum Key Distribution (QKD)~\cite{QKD_Global_2021}. However, the throughput of  mainstream implementations of QKD is limited by an intrinsically bounded Hilbert space due to the reliance on 2-dimensional polarization encoding. One way to overcome such a limitation is to use Degrees of Freedom (DoF) of light (in this work we shall take the term ``light'' to mean a single photon) that give access to a higher-dimensional Hilbert space for quantum encoding.

The helicity (polarization) and the three momentum components  represent the DoF for a photonic mode. These four fundamental DoF can manifest themselves in a host of different forms, such as time, energy, and spatial --- the Orbital Angular Momentum (OAM) being but one of several spatial forms (i.e., a transverse spatial mode). Some of these DoF are  complementary observables, e.g., time and frequency.
The OAM of light has been considered as a promising candidate for high-throughput quantum communications due to its infinite-dimensional Hilbert space. Indeed, extensive research efforts have been devoted towards the use of OAM of light in free-space entanglement distribution (e.g.,~\cite{OAM_entanglement_3km,ZiqingOAMEntDist}) and quantum communications (e.g.,~\cite{OAM_QKD_Qubit_Hybrid2017,Ziqing_OAMQKD}).

The Temporal Modes (TMs) of light, which also provide an infinite-dimensional Hilbert space for quantum encoding, open up new possibilities.
TMs are a set of overlapping, but orthogonal, broadband wave-packet modes that form a complete basis for representing an arbitrary quantum state~\cite{CompleteFramework,ThenAndNow}.
It has been pointed out that all operations necessary for quantum communications can be implemented with TMs~\cite{CompleteFramework}.
The recent advances, including controlled TM generation (e.g.,~\cite{TM_StateGen_Remote2020}), targeted TM manipulation (e.g.,~\cite{TM_ArbTrans2020}), and  high-efficiency TM-selective measurement (e.g.,~\cite{QPG2011_First,TMResolvedPhotonCounting2013,TM_Measurement2014,QPG_Experiment_OptPump_2016,QPG_Exp_2017,TMI_Exp_2018}), have greatly improved the practical feasibility of using the TMs of light in quantum communications.

Despite their potential in high-throughput quantum communications, the TMs of light have received less attention than the OAM of light over the years. It has been estimated that TM QKD is feasible over an optical fiber of $\sim\!\!100\,\text{km}$ with current technologies~\cite{QPG_Experiment_OptPump_2016}. However, the feasibility of utilizing the TMs of light in free-space quantum communications still remains unknown. In particular, it is unknown whether a TM-based system can outperform an OAM-based system in satellite-to-Earth quantum communications.
Recently, we have shown that the OAM of light can enable the theoretically-predicted key rate advantage provided by higher-dimensional QKD over a satellite-to-Earth channel~\cite{Ziqing_OAMQKD}.

In this work, we investigate the feasibility of using the TMs of light in satellite-to-Earth quantum communications. Specifically, our contributions  are as follows:
\begin{itemize}
	\item [(i)] We evaluate the detection performance for TM states, finding that TM states are more robust than OAM states against the negative propagation effects within a satellite-to-Earth  atmospheric channel.
	\item [(ii)] We determine the level of {{TM sorting performance (i.e., TM discrimination)}} that a TM-based system requires to outperform an OAM-based system in satellite-to-Earth quantum communications.
	\item [(iii)] We consider an actual application, satellite-to-Earth TM QKD, showing that the TMs of light can be superior to the OAM of light over a satellite-to-Earth channel.
    \end{itemize}
Overall, our work provides new  insights into the future development of high-dimensional single-photon satellite-to-Earth quantum communication systems. We highlight that the TMs of photons are likely the superior quantum modes to be exploited in this regard.


\vspace{-0.2cm}
\section{Temporal Modes of Light}
For a fixed polarization and transverse field distribution, a single-photon quantum state in a specific TM (denoted as $\ket{A_{n}}$) can be expressed as a superposition of single-photon states created over creation times, viz.,
	$\ket{A_{n}}=\int {d t} F_{n}(t) \hat{A}^{\dagger}(t)\ket{0}$,
where $n\!\ge\!0$ denotes the TM order,  $\hat{A}^{\dagger}(t)$ denotes the  operator that creates a photon at time $t$, $\ket{0}$ denotes the vacuum state, and $F_{n}(t)$ denotes the complex temporal amplitude of the wave packet~\cite{CompleteFramework}. Through the Fourier transform, we will find it convenient to describe the same state $\ket{A_{n}}$  as a superposition of single-photon states in different frequency modes
$	\ket{A_{n}}=\int \frac{d \omega}{2 \pi} f_{n}(\omega) \hat{a}^{\dagger}(\omega)\ket{0}$,
where $\omega$ denotes (angular) frequency, $\hat{a}^{\dagger}(\omega)$ denotes the  creation
operator at $\omega$, and $f_n(\omega)$ denotes the complex spectral amplitude of the wave packet defined as the Fourier transform of $F_{n}(t)$~\cite{CompleteFramework}.
The TMs are orthogonal with respect to a frequency integral, $\braket{A_j}{A_k}=\frac{1}{2\pi}\! \int \! d \omega f_{j}^{*}(\omega) f_{k}(\omega)=\delta_{j,k}$, where $\delta$ denotes the Kronecker delta function.
 


We choose the TM basis to be a family of Hermite-Gaussian (HG) functions, making TMs correspond to HG pulses (the terms ``pulse'' and ``single-photon wave packet'' are used interchangeably in this work)~\cite{CompleteFramework,ThenAndNow}. The complex spectral amplitude of the $n^{\text{th}}$ order TM is given by
\begin{equation}\label{Eq:HG_FD}
	f_n(\omega;\omega_0)=\mathcal{N}\, H_{n}\left(\frac{\omega-\omega_{0}}{\sigma}\right) \exp\left[{-\frac{\left(\omega-\omega_{0}\right)^{2}}{2 \sigma^2}}\right],
\end{equation}
where $\mathcal{N}$ is a normalization factor, $H_n(\cdot)$ is the $n^{\text{th}}$ order Hermite polynomial, $\sigma$ is a standard deviation (in frequency) related to the spectral width of the pulse, and $\omega_0$ is the central frequency~\cite{TM_Measurement2014}.
The Full Width at Half Maximum (FWHM) temporal pulse duration of the $0^{\text{th}}$ order TM is given by $T_0=2\sqrt{\ln2}/\sigma$, and we refer to this quantity as the \textit{pulse duration}  throughout this work.

\vspace{-0.2cm}
\section{Atmospheric Propagation}
We denote the satellite and the ground station as Alice (sender) and Bob (receiver), respectively.
The ground-station altitude is denoted as $h_0$, the satellite zenith angle at the ground station is denoted as $\theta_z$, and the satellite altitude at $\theta_z\!=\!0$ is denoted as $H$. The channel distance $L$ is given by $L\!=\!(H\!-\!h_0)/\cos\theta_z$. We denote the aperture radius at the ground station receiver as $r_a$.  We consider a Low-Earth-Orbit (LEO) satellite with a fixed satellite altitude $H\!=\!500\,\text{km}$, and set the zenith angle $\theta_z\!=\!0\degree$ for simplicity.

As is common, we assume that the solution of the wave equation is factorized in space and time so that the spatial characteristics and the spectral-temporal characteristics do not interact during propagation (valid for optical pulses of several tens of femtoseconds or longer, see e.g.,~\cite{UltrashortPGB}).
To investigate the spectral-temporal domain pulse evolution, we model the atmosphere as a lossless dispersive linear optical medium with a frequency-dependent refractive index.
The resulting chromatic dispersion imposes negative effects (e.g., broadening and distortion) on an optical pulse as it propagates, which can be modeled separately from the spatial distortion.
We set the central wavelength to $\lambda_0\!=\!1.064\,\upmu$m (corresponding to $\omega_0=2\pi\times 282\,\text{THz}$), and set the pulse duration to $T_0=200\,\text{fs}$ by adjusting $\sigma$ in Eq.~(\ref{Eq:HG_FD}).
To determine the effects of dispersion within the satellite-to-Earth channel, we follow the common practice (see e.g.,~\cite{GibbinsShortPulsePropagation1990}) of dividing the atmosphere vertically into $1\,\text{km}$ thick layers to an altitude of $100\,\text{km}$ (we treat the atmosphere above $100\,\text{km}$ as a single layer). The $q^{\text{th}}$ layer is bounded by two specific altitudes $h_{q-1}$ and $h_q$ (with $q$ ranging from 1 to 101).
The slant propagation distance in each layer is calculated as $L_q=(h_q-h_{q-1})/\cos(\theta_z)$.

We first consider the scenario where Alice sends a single-photon TM  $\ket{A_{n_t}}$ to Bob's ground station through the  atmospheric channel.
Denoting the complex spectral amplitude of the transmitted single-photon TM  $\ket{A_{n_t}}$ as $f_{n_t}(\omega;\omega_0)$, the complex spectral amplitude of the received state $\ket{A_{n_t}^L}$ is expressed as
\begin{equation}\label{EqRealPulsePropFreqDomain}
	f_{n_t}^{L}(\omega;\omega_0)=f_{n_t}(\omega;\omega_0) \exp \left[-i \sum_{q} k_{q}(\omega) L_{q}\right],
\end{equation}
where $k_{q}(\omega)=\langle n(\lambda,h_{q-1})\rangle \frac{\omega}{c}$ is the frequency-dependent propagation factor within the $q^{\text{th}}$ layer of the atmosphere~\cite{GibbinsShortPulsePropagation1990,FSLaserBook}. Note that $\langle n(\lambda,h_{q-1})\rangle$ denotes the mean refractive index of atmosphere (which depends on both wavelength $\lambda$ and altitude $h_{q-1}$ \cite{SUPP}), and $c$ denotes the speed of light in vacuum.
The propagation factor $k_{q}(\omega)$ can be expanded at the central frequency $\omega_0$ using a Taylor series expansion
\begin{equation}\label{Eq.BetaExpension}
	\begin{aligned}
		k_{q}(\omega)\!=\! k_{q}\left(\omega_{0}\right) \!+\! k_q^{'}(\omega_0)\left(\omega-\omega_{0}\right)\!+\!\frac{1}{2}k_q^{''}(\omega_0)\left(\omega-\omega_{0}\right)^{2}+\cdots,
	\end{aligned}
\end{equation}
where $k_q^{'}(\omega_0)=\left.\frac{\mathrm{d} k_{q}(\omega)}{\mathrm{d} \omega}\right|_{\omega_{0}}$
and $k_q^{''}(\omega_0)=\left. \frac{\mathrm{d}^{2} k_{q}(\omega)}{\mathrm{d} \omega^{2}}\right|_{\omega_{0}}$~\cite{FSLaserBook}.
We apply a phase factor $\exp[i\omega\sum_q {L_q}k_q^{'}(\omega_0)]$ to $f_{n_t}^{L}(\omega;\omega_0)$ in order to remove the total time-shift (i.e., group delay) relative to the original transmitted pulse.

We can encapsulate the above discussion more formally by describing the TM quantum state evolution  as
\begin{equation}\label{Eq.TMEvo}
	\ket{A_{n_t}^L}=U_{\text{disp}}(L)\ket{A_{n_t}}=\sum_{n} c_{n, {n_t}}(L)\ket{A_n},
\end{equation}
where $U_{\text{disp}}(L)$ is a unitary operator that describes the evolution of a single-photon TM including all terms of Eq.~(\ref{Eq.BetaExpension}), and $c_{n, n_{t}}(L)=\left\langle A_{n}\left|U_{\text {disp}}(L)\right| A_{n_t}\right\rangle$ is the coefficient for expanding $\ket{A_{n_t}^L}$ in the TM basis. After a non-dispersive propagation  we have $c_{n,n_t}\!=\!\delta_{n,n_t}$ due to the orthogonality of TM states. In reality, however, dispersion-induced effects introduce crosstalk. At the receiver, $\ket{A_{n_t}^L}$ is generally a superposition of $\ket{A_n}$, and thus it is no longer orthogonal to any TM state. The crosstalk probability that an original TM $\ket{A_{n_t}}$ scatters into another TM $\ket{A_n}$ is calculated as $P_{\text{Ch}}(n|{n_t})=|c_{n,n_t}|^2$.

The concept of dispersion compensation has been widely implemented in fields such as classical communications (see e.g.,~\cite{FiberOpticsBook}) and laser ranging (see e.g.,~\cite{fsPulseToA2010}).
In this work we further consider the improvements provided by dispersion compensation at the single-photon level (at the receiver).
Known as the second-order dispersion, the Group Delay Dispersion (GDD) contributes heavily to the overall dispersion at optical frequencies within the atmosphere (see discussions in e.g.,~\cite{fsPulseToA2010}). We only compensate for GDD when applying dispersion compensation.
The complex spectral amplitude of the dispersion-compensated received state $\ket{\tilde{A}_{n_t}^L}$ is given by
\begin{equation}\label{Eq.RealPulsePropFreqDomain_GVDC}
	\begin{aligned}
		\tilde{f}_{n_t}^{L}(\omega;\omega_0)=f_{n_t}^{L}(\omega;\omega_0) \exp\left(i\frac{1}{2}(\omega-\omega_0)^2\sum_{q}k^{''}_q(\omega_0)L_q\right).
	\end{aligned}
\end{equation}
Formally, the dispersion-compensated version of Eq.~(\ref{Eq.TMEvo}) is obtained by replacing $\ket{A_{n_t}^L}$ with $\ket{\tilde{A}_{n_t}^{L}} = U_{\text{comp}}\ket{A_{n_t}^{L}}$,
where $U_{\text{comp}}$ denotes the operation of dispersion compensation. Note,  $U_{\text{comp}}$  does not fully compensate the effects of the dispersion as the  higher order terms of Eq.~(\ref{Eq.BetaExpension}) are not corrected for (i.e., $U_{\text{comp}} \ne U^{-1}_{\text{disp}})$. As such, dispersion compensation in our simulations will not be perfect --- only  GDD is corrected for.

Recall, we will compare our results on the use of TMs relative to the use of OAM single-photon states. The quantum state evolution of an OAM state is determined by the transverse field evolution in the spatial domain. To investigate the transverse field evolution in the spatial domain, we model the atmosphere as a  turbulent (random) medium that imposes random amplitude and phase distortions on the transverse optical field as it propagates.
We set the frequency of the transverse optical field to the central frequency $\omega_0$, and  set the beam-waist radius to $15\,\text{cm}$.
We set the sea-level turbulence strength to $9.6 \times 10^{-14}\, \text{m}^{-2 / 3}$, and the root-mean-square wind speed to $21\,\text{m/s}$. We set the outer scale  and the inner scale of the atmospheric turbulence to $5\,\text{m}$ and $1\,\text{cm}$, respectively. Explanations for the choice of the above parameters are given in~\cite{Ziqing_OAMQKD}, and details on how we simulate the pulse and transverse field evolutions are provided in \cite{SUPP}.

\vspace{-0.3cm}
\section{Temporal Mode Discrimination}\label{Sec.TMSorter}
\begin{figure}[!hbtp]
	\centering
	\vspace{-0.2cm}
	\includegraphics[width=\columnwidth]{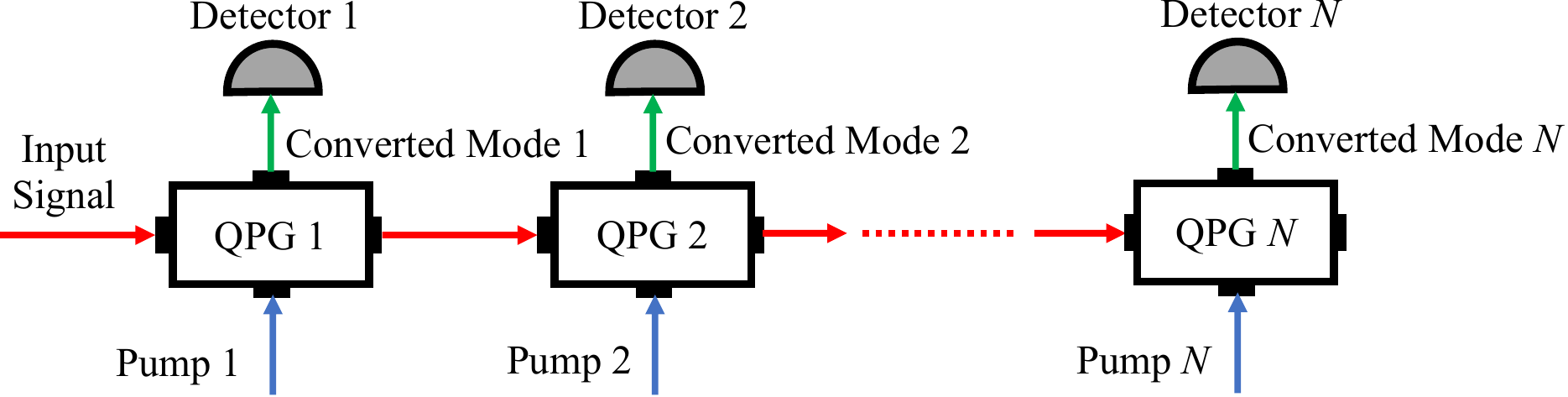}
	\vspace{-0.6cm}
	\caption{The TM sorter realized by a chain of QPGs.}
	\label{Fig.TMSorter}
	\vspace{-0.25cm}
\end{figure}

A promising device to achieve a TM-selective measurement is the Quantum Pulse Gate (QPG) based on dispersion-engineered Frequency Conversion (FC) inside a nonlinear optical waveguide~\cite{QPG2011_First}.
Commonly implemented with three-wave mixing, an ideal QPG selects a certain TM state (determined by the spectral-temporal shape of the bright pump pulse that drives the FC) from the input signal and frequency-converts it to an output TM state with an efficiency of 1. Ideally, all other orthogonal TM states are completely transmitted~\cite{CompleteFramework,TailoringNL_TM_Rev2018}. The frequency-converted output signal can be separated (e.g., using a dichroic mirror) and then be detected with a photodetector to achieve a TM-selective measurement. By appropriately shaping the pump pulse, the TM state selected by a QPG can be reconfigured on the fly~\cite{TailoringNL_TM_Rev2018}.
It has been proposed that a TM sorter (that discriminates amongst a set of $N$ orthogonal TM states and sends them to their corresponding detectors) can be realized by a chain of $N$ QPGs (see e.g.,~\cite{TMResolvedPhotonCounting2013,CompleteFramework}).
The TM sorter we consider is illustrated in Fig.~\ref{Fig.TMSorter}. Specifically, all the QPGs share the identical nonlinear waveguide design and are driven by their corresponding pump pulses. Each of the $N$ QPGs serves to select one of the $N$ orthogonal TM states.

Despite the great potential of QPGs in TM-based quantum communications, the implementation of a near-ideal QPG is significantly challenging. Denoting the conversion efficiencies of different input TM states $j$ achieved with a fixed QPG pump pulse $k$ ($1\le j,\,k\le N$) as $\{\eta_{kj}\}$, an important performance metric of a QPG is given by the separability $S_k=\frac{\eta_{k k}}{\sum_{j=1}^{N} \eta_{k j}}$~\cite{QPG_Experiment_OptPump_2016}.
Note that, for single-photon inputs  the conversion efficiency $ \eta_{k j}$ can be understood as a probability (i.e., the probability that a photon in TM state $j$ gets converted by a QPG driven by pump $k$).
An ideal TM-selective QPG should achieve $\eta_{k j}=\delta_{k,j}$, leading to $S_k=1$. However, for a single QPG there exists a fundamental limit that enforces a trade-off between $\eta_{k k}$ and $S_k$. This fundamental limit exists because the effects of time-ordering lead to  non-negligible $ \eta_{k j}$ ($j\ne k$) at large $\eta_{k k}$ values (see discussions in e.g.,~\cite{CompleteFramework,QPG_Exp_2017,TMI_Exp_2018}). The TM discrimination (i.e., TM sorting performance) is determined by $\{S_k\}$.

We take into account the effects of imperfect TM sorting.
For simplicity, we assume that the $k^{\text{th}}$ QPG constituting the TM sorter (i) selects and converts  TM state $k$ with efficiency $\eta_{kk}=0.9$, (ii) selects and converts other $N-1$ orthogonal TM states with the same efficiency $ \eta_{k j}\,(j\ne k)$ , and (iii) passes the remaining signal to the next QPG without any modifications. We further assume that all QPGs achieve the same performance, and thus we denote $\eta_{kk}=\eta_0$, $ \eta_{k j}\,(j\ne k) = \eta_1$, and $S_k=S$.
We term $\eta_0$ and $\eta_1$ as the QPG selection factor and the QPG  error factor, respectively.
Specifically, we set $\eta_1$ ranging from 0 to $\eta_{0}$ in order to simulate different levels of imperfectness in  TM-selective measurements. Such a range of $\eta_1$ translates into $S$ ranging from 1 (best) to $1/N$ (worst). A higher $\eta_{1}$ leads to a lower $S$, indicating a worse TM sorting performance.
Assuming a photon in TM state $j$ enters the TM sorter, the probability that this photon is measured to be in TM state $k$ (due to a frequency conversion at QPG $k$) is calculated as
\begin{equation}\label{Eq.ProbTransDemux}
P_{\text{Srt}}(k|j)=
\begin{cases}
	(1-\eta_{1})^{k-2}(1-\eta_{0})\eta_{1},&k>j,\\
	(1-\eta_{1})^{k-1}\eta_{0},&k=j,\\
	(1-\eta_{1})^{k-1} \eta_{1},&k<j.\\
\end{cases}
\end{equation}

\vspace{-0.3cm}
\section{Detection Performance}\label{Sec.StateDetection}
We now investigate the detection performance (the error probability conditioned on detection) for TM states, achieved with imperfect TM sorting, over the satellite-to-Earth channel.  We consider a TM-based system that uses a $d$-dimensional encoding subspace $\mathcal{H}_d$ spanned by $d$ mutually orthogonal TM states chosen from the TM basis.
We assume that each QPG constituting the TM sorter in Fig.~\ref{Fig.TMSorter}  only selects from the $d$ TM states to be sorted.
Under this assumption, any state entering the TM sorter is projected onto the encoding subspace $\mathcal{H}_d$ (note that, such a projection is also necessary for OAM-based systems~\cite{Ziqing_OAMQKD}). The losses outside $\mathcal{H}_d$ (due to dispersion-induced crosstalk) are accommodated by collecting all other possibilities together into one projection operator.

For conciseness we index the $d$ TM states in ascending order. We assume that the $d$ QPGs constituting the TM sorter in Fig.~\ref{Fig.TMSorter} have their pumps arranged in the same order. For example, when TM states $\{\ket{A_0}, \ket{A_3}\}$ are used to construct a 2-dimensional encoding subspace $\mathcal{H}_2$, $\ket{A_0}$ ($\ket{A_3}$) is addressed as the $1^{\text{st}}$ ($2^{\text{nd}}$) TM state. To sort these two states, the TM sorter consists of two QPGs with the $1^{\text{st}}$ ($2^{\text{nd}}$) QPG addressing $\ket{A_0}$ ($\ket{A_3}$).
The probability that a photon sent in the $s^{\text{th}}$ ($1\le s \le d$) TM   is measured to be in the $m^{\text{th}}$ ($1\le m \le d$) TM  is calculated as
\begin{equation}\label{Eq.ProbTransTotal}
	P_{\text{Tot}}(m|s)=\frac{\sum_{r = 1}^{d}  P_{\text{Srt}}(m|r)P_{\text{Ch}}(r|{s})}{\mathcal{T}_{s}},
\end{equation}
where $P_{\text{Ch}}(r|{s})$ denotes the crosstalk probability that the $s^{\text{th}}$ TM  scatters into the $r^{\text{th}}$ ($1 \le r \le d$) TM  after a dispersive propagation, and
	$\mathcal{T}_{s}=\sum_{m = 1}^{d} {\sum_{r = 1}^{d}  P_{\text{Srt}}(m|r)P_{\text{Ch}}(r|{s})}$
is required for normalization. Note that $\mathcal{T}_{s}$ quantifies the photon survival fraction of photons sent in the $s^{\text{th}}$ TM, and the overall photon survival fraction is given by $\mathcal{T} = \frac{1}{d}\sum_{s=1}^{d}\mathcal{T}_{s}$. The error probability is  calculated as
\begin{equation}\label{Eq.Pe}
	P_e = \frac{1}{d} \sum_{\begin{subarray}{c} m, s\\ m \ne s \end{subarray}} P_{\text{Tot}}(m|s).
\end{equation}

For  comparison, we use the same formalism above to evaluate the detection performance for OAM states. The encoding subspace $\mathcal{H}_d$ of an OAM-based system is spanned by $d$  OAM basis states (i.e., OAM eigenstates with Laguerre-Gaussian transverse field distribution). The turbulence-induced crosstalk among OAM states is determined by the transverse field evolution in the spatial domain. For OAM states the crosstalk probability $P_{\text{Ch}}$  needs to be first averaged over realizations of atmospheric turbulence (see~\cite{Ziqing_OAMQKD} and the references therein for more details).
Note, in this work we will always assume the OAM detector itself is, in effect, noiseless and causes no error detection by itself (we term such a detector  as a `perfect sorter')
 due to the existence of well-developed tools (see e.g.,~\cite{OAMSorter2013}).


We consider dimensions ranging from $d=2$ to $d=9$, and we assume TM  basis states with TM orders ranging from 0 to 8 (OAM basis states with OAM numbers ranging from -4 to 4) are available to construct the encoding subspace $\mathcal{H}_d$ for the TM-based (OAM-based) system. For each dimension $d$ the encoding subspace $\mathcal{H}_d$ is optimized to minimize the error probability $P_e$.

\begin{figure}[!hbtp]
	\centering
	\includegraphics[width=\columnwidth]{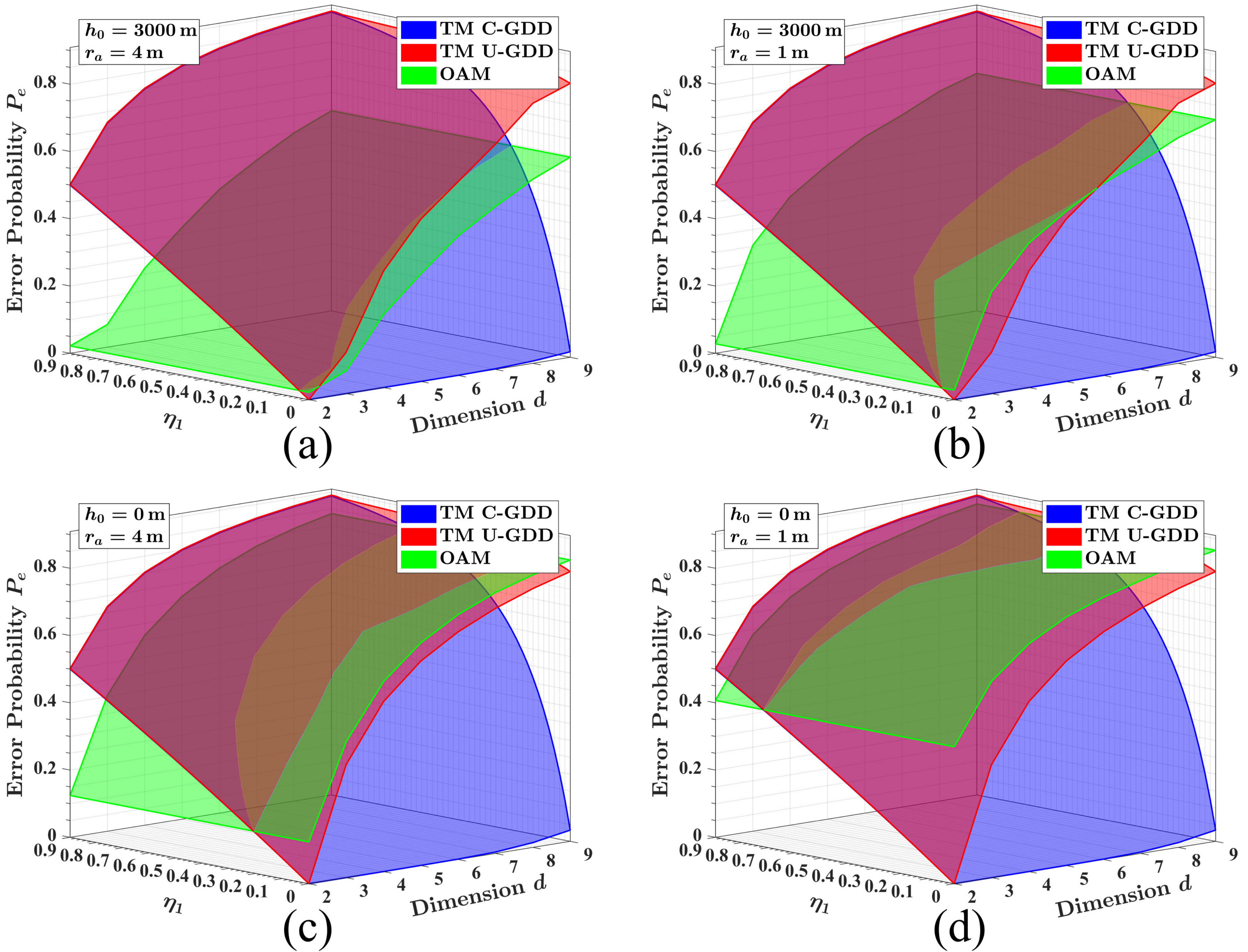}
	\vspace{-0.65cm}
	\caption{Error probabilities $P_e$ for TM states and OAM states, under (a)(b) $h_0=3000\,\text{m}$ and (c)(d) $h_0=0\,\text{m}$, achieved with (a)(c) $r_a=4\,\text{m}$ and (b)(d) $r_a=1\,\text{m}$. Results are plotted against $\eta_1$ and dimension $d$. The error probabilities for TM states achieved with corrected GDD compensation (C-GDD) and with uncorrected GDD (U-GDD) are both plotted. The error probability for OAM states is plotted independent of $\eta_1$ as a benchmark.}
	\label{Fig.Pe}
	\vspace{-0.6cm}
\end{figure}

In Fig.~\ref{Fig.Pe} we plot the error probabilities $P_e$ for TM states and OAM states, achieved with different $r_a$ values, under different $h_0$ values, against $\eta_1$ and dimension $d$.
From all sub-figures we observe that the GDD  compensation significantly reduces the error probability for TM states.
Specifically, given a certain dimension $d$, the error probability for TM states (achieved with the help of dispersion compensation) is much lower than the error probability for OAM states when $\eta_1$ is small (i.e., when TM sorting performance is good).
However, when $\eta_1$ exceeds a certain threshold the error probability for TM states surpasses the error probability for OAM states.

\vspace{-0.25cm}
\section{Quantum Key Distribution}
Finally, we consider the actual application of a quantum information protocol, namely high-dimensional QKD, to demonstrate the potential usefulness of TM states  (relative to OAM states) in the satellite-to-Earth channel.
Specifically, we investigate  TM-QKD performance achieved with imperfect TM sorting 
 and OAM-QKD performance achieved with perfect sorting. By QKD performance we mean the secret key rate $K$ in {bits per photon} (or more exactly {bits per \emph{sent} photon}).

A QKD protocol involves state Preparation and Measurement (P\&M) in multiple bases (we only consider the P\&M paradigm for QKD). Each of these bases is called a Mutually Unbiased Basis (MUB). For performance evaluation we consider a $d$-dimensional QKD protocol using $d+1$ MUBs. The first MUB is the standard basis (TM basis or OAM basis) whose states span the $d$-dimensional encoding subspace $\mathcal{H}_d$. Every other MUB consists of $d$ orthogonal states that can each be expressed as a superposition of the standard basis states (see~\cite{Ziqing_OAMQKD} and the references therein for more details regarding the mathematical formalism of MUBs).
For simplicity, we assume that all QPGs constituting the TM sorter in Fig.~\ref{Fig.TMSorter} are properly configured to achieve the same behaviors (described in Section~\ref{Sec.TMSorter}) when sorting  $d$ orthogonal states in any of the total $d+1$ MUBs. Note that a QPG can indeed select arbitrary superpositions of TM states with appropriately shaped pump pulses (see e.g.,~\cite{CompleteFramework,QPG_Experiment_OptPump_2016,TailoringNL_TM_Rev2018}).

Using the same formalism in Section~\ref{Sec.StateDetection} we can evaluate the error probability $P_e^{\beta}$ and the overall photon survival fraction $\mathcal{T}^{\beta}$ in the $\beta^{\text{th}}$ MUB (it is clear that $P_e^{1}=P_e$ and $\mathcal{T}^{1}=\mathcal{T}$). Note that $\{s,\,r,\,m\}$ of  Section~\ref{Sec.StateDetection} now denote the indices of states in a specific MUB. The average error rate (i.e., the error probability averaged over all MUBs) is then calculated as
	$Q = \frac{1}{d+1} \sum_{\beta=1}^{d+1} P_e^{\beta}$.
The key rate per photon used in key generation
can be written as a function of $Q$~\cite{Qudit_SecurityProof2012},
\begin{equation}\label{Eq:KeyRate_d+1MUB}
	\begin{aligned}
		K_1=& \log _{2} d+\frac{d+1}{d} Q \log _{2}\left(\frac{Q}{d(d-1)}\right) \\ &+\left(1-\frac{d+1}{d} Q\right) \log _{2}\left(1-\frac{d+1}{d} Q\right).
	\end{aligned}
\end{equation}

To link $K_1$ to $K$ we first calculate the average photon survival fraction (i.e., the overall photon survival fraction averaged over all MUBs) as $\mathcal{T}_{\text{avg}} = \frac{1}{d+1} \sum_{\beta=1}^{d+1} \mathcal{T}^{\beta}$.
For practical reasons we further assume that TM (OAM) QKD only uses the photons in the fundamental Gaussian spatial mode ($0^\text{th}$ order TM). Due to the resulting crosstalk in the spatial domain (spectral-temporal domain), the atmospheric turbulence (dispersion) causes extra photon loss in TM (OAM) QKD. We define the spatial mode (temporal mode) mismatch coefficient $C_{\text{SMM(TMM)}}$ to quantify such a photon loss. Specifically, the spatial mode mismatch coefficient is given by
$C_{\text{SMM}} = \left\langle{\left|\int\!\!\int\psi_0^{*}(x,y,L)A(x,y)\psi_0^{\text{turb}}(x,y,L)\,\text{d}x\,\text{d}y\right|^2}\right\rangle$
where $\psi_0(x,y,L)$ ($\psi_0^{\text{turb}}(x,y,L)$) denotes the fundamental Gaussian spatial mode after propagating through a vacuum (turbulent) channel, $A(x,y)$ denotes the receiver aperture function, and $\ensavg{\cdots}$ denotes an ensemble average over different realizations of the turbulent atmospheric channel. The temporal mode mismatch coefficient is given by $C_{\text{TMM}} = \left|\frac{1}{2\pi}\int f_0^{*}(\omega;\omega_0)f_0^{L}(\omega;\omega_0)\,\text{d}\omega\right|^2$.
After dispersion compensation, the temporal mode mismatch coefficient is given by
$C_{\text{TMM}}^{'} = |\frac{1}{2\pi}\int f_0^{*}(\omega;\omega_0)\tilde{f}_{0}^{L}(\omega;\omega_0) \,\text{d}\omega|^2$.
The secret key rate $K_{\text{TM(OAM)}}$ of TM (OAM) QKD is then given by
\begin{equation}\label{Eq:FinalKey}
	\begin{aligned}
	K_{\text{TM(OAM)}} &= C_{\text{SMM(TMM)}}\cdot\mathcal{T}_{\text{avg}}^{\text{TM(OAM)}}\cdot K_{1}^{\text{TM(OAM)}}. \\
	\end{aligned}
\end{equation}

To compare $K_{\text{TM}}$ and $K_{\text{OAM}}$, we adopt all settings (including dimension range, range of available basis states, and simulation parameters) as described in Section~\ref{Sec.StateDetection}.
For OAM-QKD we further assume that the quantum channel information is available and a \textit{quantum channel conjugation} is applied to the received OAM states (see~\cite{Ziqing_OAMQKD} for details).
For each dimension $d$ the encoding subspace $\mathcal{H}_d$ is optimized to maximize the secret key rate $K$.

\begin{figure}[!hbtp]
	\centering
	\vspace{-0.4cm}
	\includegraphics[width=\columnwidth]{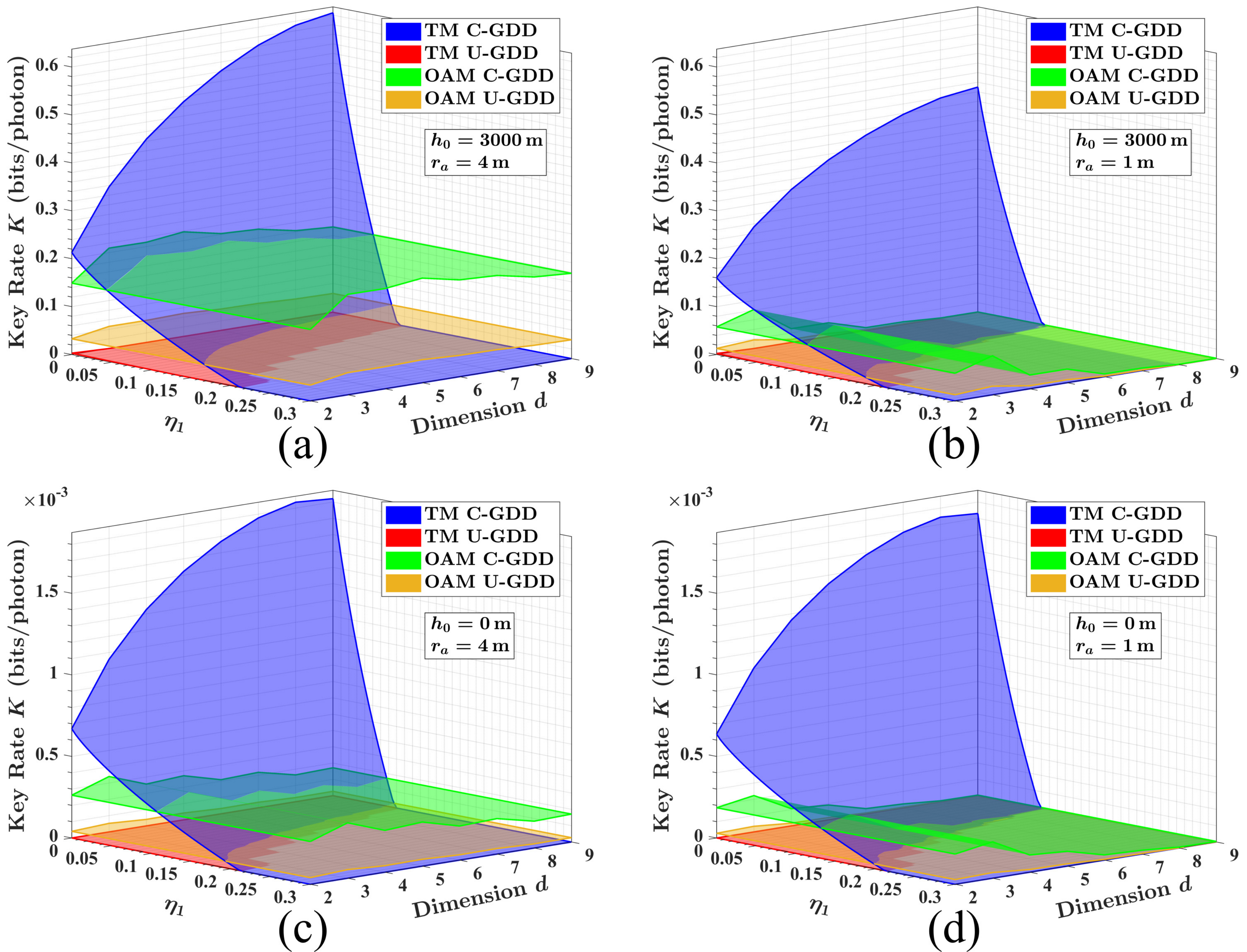}
	\vspace{-0.65cm}
	\caption{Secret key rates $K$ for TM-QKD and OAM-QKD, under (a)(b) $h_0=3000\,\text{m}$ and (c)(d) $h_0=0\,\text{m}$, achieved with (a)(c) $r_a=4\,\text{m}$ and (b)(d) $r_a=1\,\text{m}$. Results are plotted against $\eta_1$ and dimension $d$. Secret key rates achieved with  
 corrected GDD compensation (C-GDD) and with uncorrected GDD (U-GDD)  
 are plotted. $K_{\text{OAM}}$ is plotted independent of $\eta_1$ as a benchmark. Here we only plot the results within the range $0\!\le\!\eta_1\!\le\!0.3$ since $K_{\text{TM}}$ drops to zero  before $\eta_1$ reaches 0.3. Note that  $K_{\text{OAM}}$ is always positive (although can be very close to zero) due to the use of quantum channel conjugation. Also, note that the increase in $K_{\text{OAM}}$ with dimension $d$ is hardly noticeable  (for example, maximum $K_{\text{OAM}}$ in subfigure (a) is achieved when $d=5$).}
	\label{Fig.SKR}
	\vspace{-0.3cm}
\end{figure}
In Fig.~\ref{Fig.SKR} we plot the secret key rates $K$ of TM-QKD and OAM-QKD, achieved with different $r_a$ and $h_0$ values, against $\eta_1$ and dimension $d$. From all sub-figures we observe that  TM-QKD (i) outperforms OAM-QKD, and (ii) enables the high-dimensional key rate advantage better than OAM-QKD over the satellite-to-Earth channel as long as the TM sorting performance is beyond a certain level. However, as expected, TM-QKD is eventually outperformed by OAM-QKD as the sorting performance degrades. Note that in this figure we have also considered GDD compensation as applied to the OAM-QKD. This is because when considering key \emph{rates} the QKD performance is ultimately linked to the pulse rate (and therefore the shape of the temporal pulse). C-GDD allows for more of the initial pulse to remain within the $0^{\text{th}}$ order TM pulse (the pulse we consider the initial OAM state to be in).

\section{Conclusions}
\vspace{-0.05cm}
In this work, we numerically investigated the feasibility of utilizing the TMs of light in high-throughput satellite-to-Earth quantum communications. Both the detection performance and the QKD performance were evaluated over the satellite-to-Earth channel.
Our results indicate that using the TMs of light is preferable to using the OAM of light in satellite-to-Earth quantum communications as long as the TM sorting performance is beyond a certain level, typically in the vicinity of 10\% QPG error factor.
We thank Benjamin Burnett for useful discussions.

\vspace{-0.35cm}
\bibliographystyle{IEEEtran}

\begin{thebibliography}{10}
	\providecommand{\url}[1]{#1}
	\csname url@samestyle\endcsname
	\providecommand{\newblock}{\relax}
	\providecommand{\bibinfo}[2]{#2}
	\providecommand{\BIBentrySTDinterwordspacing}{\spaceskip=0pt\relax}
	\providecommand{\BIBentryALTinterwordstretchfactor}{4}
	\providecommand{\BIBentryALTinterwordspacing}{\spaceskip=\fontdimen2\font plus
		\BIBentryALTinterwordstretchfactor\fontdimen3\font minus
		\fontdimen4\font\relax}
	\providecommand{\BIBdecl}{\relax}
	\BIBdecl
	\vspace{-0.05cm}
	\bibitem{QKD_Global_2021}
	Y.-A. Chen \emph{et~al.}, ``An integrated space-to-ground quantum communication
	network over 4600 kilometres,'' \emph{Nature}, vol. 589, no. 7841, pp.
	214--219, 2021.
	
	\bibitem{OAM_entanglement_3km}
	M.~Krenn \emph{et~al.}, ``Twisted photon entanglement through turbulent air
	across {V}ienna,'' \emph{Proc. Natl. Acad. Sci. U.S.A.}, vol. 112, no.~46,
	pp. 14\,197--14\,201, 2015.
	
	\bibitem{ZiqingOAMEntDist}
	Z.~{Wang}, R.~{Malaney}, and J.~{Green}, ``Satellite-based entanglement
	distribution using orbital angular momentum of light,'' in \emph{Proc. IEEE
		Int. Conf. Commun. Workshops (ICC Wkshps)}, 2020.
	
	\bibitem{OAM_QKD_Qubit_Hybrid2017}
	A.~Sit \emph{et~al.}, ``High-dimensional intracity quantum cryptography with
	structured photons,'' \emph{Optica}, vol.~4, no.~9, pp. 1006--1010, 2017.
	
	\bibitem{Ziqing_OAMQKD}
	Z.~Wang, R.~Malaney, and B.~Burnett, ``Satellite-to-{E}arth quantum key
	distribution via orbital angular momentum,'' \emph{Phys. Rev. Applied},
	vol.~14, p. 064031, 2020.
	
	\bibitem{CompleteFramework}
	B.~Brecht \emph{et~al.}, ``Photon temporal modes: A complete framework for
	quantum information science,'' \emph{Phys. Rev. X}, vol.~5, p. 041017, 2015.
	
	\bibitem{ThenAndNow}
	M.~G. Raymer and I.~A. Walmsley, ``Temporal modes in quantum optics: then and now,'' \emph{Physica Scripta}, vol.~95, no.~6, p. 064002, 2020.
	
	\bibitem{TM_StateGen_Remote2020}
	V.~Ansari \emph{et~al.}, ``Remotely projecting states of photonic temporal
	modes,'' \emph{Opt. Express}, vol.~28, no.~19, pp. 28\,295--28\,305, 2020.
	
	\bibitem{TM_ArbTrans2020}
	J.~Ashby \emph{et~al.}, ``Temporal mode transformations by sequential time and
	frequency phase modulation for applications in quantum information science,''
	\emph{Opt. Express}, vol.~28, no.~25, pp. 38\,376--38\,389, 2020.
	
	\bibitem{QPG2011_First}
	A.~Eckstein, B.~Brecht, and C.~Silberhorn, ``A quantum pulse gate based on
	spectrally engineered sum frequency generation,'' \emph{Opt. Express},
	vol.~19, no.~15, pp. 13\,770--13\,778, 2011.

	\bibitem{TMResolvedPhotonCounting2013}
	Y.-P. Huang and P.~Kumar, ``Mode-resolved photon counting via cascaded quantum frequency conversion,'' \emph{Opt. Lett.}, vol.~38, no.~4, pp. 468--470, 2013.

	\bibitem{TM_Measurement2014}
	B.~Brecht \emph{et~al.}, ``Demonstration of coherent time-frequency Schmidt mode selection using dispersion-engineered frequency conversion,'' \emph{Phys. Rev. A}, vol.~90, p. 030302, 2014.
	
	\bibitem{QPG_Experiment_OptPump_2016}
	P.~Manurkar \emph{et~al.}, ``Multidimensional mode-separable frequency
	conversion for high-speed quantum communication,'' \emph{Optica}, vol.~3,
	no.~12, pp. 1300--1307, 2016.
	
	\bibitem{QPG_Exp_2017}
	D.~V. Reddy and M.~G. Raymer, ``Engineering temporal-mode-selective frequency
	conversion in nonlinear optical waveguides: from theory to experiment,''
	\emph{Opt. Express}, vol.~25, no.~11, pp. 12\,952--12\,966, 2017.
	
	\bibitem{TMI_Exp_2018}
	D.~V. Reddy and M.~G. Raymer, ``High-selectivity quantum pulse gating of
	photonic temporal modes using all-optical ramsey interferometry,''
	\emph{Optica}, vol.~5, no.~4, pp. 423--428, 2018.
	
	\bibitem{UltrashortPGB}
	M.~A.~Porras, ``Ultrashort pulsed Gaussian light beams,'' \emph{Phys. Rev. E}, vol.~58, pp. 1086--1093, 1998.
		
	\bibitem{GibbinsShortPulsePropagation1990}
	C.~J.~Gibbins, ``Propagation of very short pulses through the absorptive and
	dispersive atmosphere,'' \emph{IEE Proceedings H (Microwaves, Antennas and Propagation)}, vol. 137, no.~5, pp. 304--310(6), 1990.
	
	\bibitem{FSLaserBook}
	C.~Rulli\`ere, \emph{Femtosecond Laser Pulses: Principles and
		Experiments}, 2nd~ed. Springer, New York, NY,
	2005.
	
	\bibitem{SUPP}
	See \emph{Supplementary Material} at the end of this paper.
	
	\bibitem{FiberOpticsBook}
	G.~P. Agrawal, \emph{Fiber‐Optic Communication Systems}, 4th~ed. John Wiley \& Sons, Inc., 2010.
	
	\bibitem{fsPulseToA2010}
	J.~Lee \emph{et~al.}, ``Time-of-flight measurement with femtosecond light
	pulses,'' \emph{Nat. Photonics}, vol.~4, no.~10, pp. 716--720, 2010.

	\bibitem{TailoringNL_TM_Rev2018}
	V.\,Ansari\,\,\emph{et\,al.},\,``Tailoring\,\,nonlinear\,\,processes\,\,for\,\,quantum\,\,optics\,with pulsed\,temporal-mode\,\,encodings,''\,\emph{Optica},\,\!vol.\,\!5,\,\!no.\,\!5,\,pp.\,534--550,\,2018.
	
	\bibitem{OAMSorter2013}
	M.~Mirhosseini \emph{et~al.}, ``Efficient separation of the orbital angular
	momentum eigenstates of light,'' \emph{Nat. Commun.}, vol.~4, no.~1, p. 2781,\,2013.
	
	\bibitem{Qudit_SecurityProof2012}
	A.~Ferenczi and N.~L\"utkenhaus, ``Symmetries in quantum key distribution and
	the connection between optimal attacks and optimal cloning,'' \emph{Phys.
		Rev. A}, vol.~85, p. 052310, 2012.
	
\end{thebibliography}

\begin{thebibliography}{1}
	\makeatletter
	\addtocounter{\@listctr}{24}
	\makeatother
	\providecommand{\url}[1]{#1}
	\csname url@samestyle\endcsname
	\providecommand{\newblock}{\relax}
	\providecommand{\bibinfo}[2]{#2}
	\providecommand{\BIBentrySTDinterwordspacing}{\spaceskip=0pt\relax}
	\providecommand{\BIBentryALTinterwordstretchfactor}{4}
	\providecommand{\BIBentryALTinterwordspacing}{\spaceskip=\fontdimen2\font plus
		\BIBentryALTinterwordstretchfactor\fontdimen3\font minus
		\fontdimen4\font\relax}
	\providecommand{\BIBforeignlanguage}[2]{{%
			\expandafter\ifx\csname l@#1\endcsname\relax
			\typeout{** WARNING: IEEEtran.bst: No hyphenation pattern has been}%
			\typeout{** loaded for the language `#1'. Using the pattern for}%
			\typeout{** the default language instead.}%
			\else
			\language=\csname l@#1\endcsname
			\fi
			#2}}
	\providecommand{\BIBdecl}{\relax}
	\BIBdecl
	
	\bibitem{ShortPulseEnergyDensity2014_Apndx}
	V.~Banakh and I.~N. Smalikho, ``Fluctuations of energy density of short-pulse
	optical radiation in the turbulent atmosphere,'' \emph{Opt. Express},
	vol.~22, no.~19, pp. 22\,285--22\,297, 2014.
	
	\bibitem{book_Apndx}
	L.~C. Andrews and R.~L. Phillips, \emph{Laser Beam Propagation through Random
		Media, Second Edition}.\hskip 1em plus 0.5em minus 0.4em\relax SPIE,
	Bellingham, WA, 2005.
	
	\bibitem{ITU_Standard_Atmos_Apndx}
	\BIBentryALTinterwordspacing
	\emph{Reference Standard Atmospheres}, ITU Radiocommunication Sector (ITU-R)
	Std. ITU-R P.835-6, 2017.  \url{https://www.itu.int/rec/R-REC-P.835/en}
	\BIBentrySTDinterwordspacing
	
	\bibitem{SimAP_Apndx}
	J.~D. Schmidt, \emph{Numerical Simulation of Optical Wave Propagation with
		Examples in MATLAB}.\hskip 1em plus 0.5em minus 0.4em\relax SPIE, Bellingham,
	WA, 2010.
	
	\bibitem{EduardoEnhanced2021_Apdnx}
	\BIBentryALTinterwordspacing
	E.~Villaseñor \emph{et~al.}, ``Enhanced uplink quantum communication with
	satellites via downlink channels,'' \emph{arXiv: 2102.01853 [quant-ph]},
	2021.  \url{https://arxiv.org/abs/2102.01853}
	\BIBentrySTDinterwordspacing
	
	\bibitem{ShapiroFarField_Apdnx}
	\BIBentryALTinterwordspacing
	W.~He \emph{et~al.}, ``Performance analysis of free-space quantum key
	distribution using multiple spatial modes,'' \emph{arXiv: 2105.01858
		[quant-ph]}, 2021.  \url{https://arxiv.org/abs/2105.01858}
	\BIBentrySTDinterwordspacing
	
\end{thebibliography}


\clearpage

\onecolumn
\LARGE{\begin{center} {Temporal Modes of Light in Satellite-to-Earth Quantum Communications: Supplementary Material}\end{center}}
\vspace{0.2cm}
\large{\begin{center}Ziqing Wang, Robert Malaney, and Ryan Aguinaldo\end{center}}
\vspace{0.2cm}
\normalsize

\setcounter{equation}{0}
\renewcommand{\theequation}{A\arabic{equation}}

\begin{multicols*}{2}
\section{Evolution within Atmosphere}\label{Appendix:DisperiveProp}
When studying the spectral-temporal domain pulse evolution we follow the common practice to neglect the turbulent fluctuations of the refractive index (note that the dependence of turbulent fluctuations of the refractive index on the frequency can be neglected in the dispersive atmosphere, see e.g.,~\citeApndx{ShortPulseEnergyDensity2014_Apndx}). We also neglect the transverse spatial dependence of the mean refractive index.
The mean refractive index profile of the atmosphere is given by~\citeApndx{book_Apndx}
\begin{equation}\label{Eq.RefractiveIndex}
	\ensavg{n(\lambda,{h})}=1+77.6 \times 10^{-6}\left(1+7.52 \times 10^{-3} \lambda^{-2}\right) \frac{\ensavg{P({h})}}{\ensavg{T(h)}},
\end{equation}
where $h$ denotes altitude above sea level, $\lambda$ is the optical wavelength in $\upmu$m, $\ensavg{P(h)}$ is the mean pressure in millibars, and $\ensavg{T(h)}$ is the mean temperature in Kelvin. The mean pressure $\ensavg{P(h)}$ and the mean temperature $\ensavg{T(h)}$ are both functions of altitude $h$, and we set these quantities in accord with \textit{ITU-R P.835-6: Reference Standard Atmospheres}~\citeApndx{ITU_Standard_Atmos_Apndx}. We assume $\ensavg{n(\lambda,{h})}\!\!=\!\!1$ for all considered $\lambda$ values at $h\!\ge\!100\,\text{km}$. Using this relation, the effects of dispersion within the satellite-to-Earth channel are as described in the main text.


To investigate the spatial domain transverse field evolution, we follow the common practice and model the atmosphere as a  turbulent (random) medium with random inhomogeneities (turbulent eddies) of different size scales~\citeApndx{book_Apndx}.
These turbulent eddies give rise to small random refractive index fluctuations, causing continuous phase modulations on the optical field. This leads to random refraction effects, imposing amplitude and phase distortions on the transverse optical field as it propagates through the atmospheric channel. The family of eddies bounded above by the outer scale $L_\text{outer}$ and below by the inner scale $l_\text{inner}$ form the inertial subrange~\citeApndx{book_Apndx}.

We assume that the transverse field evolutions at all considered frequencies are the same as the transverse field evolution at the central frequency $\omega_0$. Indeed, we find that the maximum difference (between values at the central frequency and the maximum/minimum frequency) in both the scintillation index and the Fried parameter is only $\!\sim\!3\%$ (see~\cite{Ziqing_OAMQKD} and the references therein for more details on these parameters).
We also follow the common practice to neglect the frequency dependence of the refractive index~\citeApndx{book_Apndx}. Specifically, the atmospheric turbulence is assumed to satisfy
\begin{equation}\label{}
	\langle n(\mathbf{R}) \rangle\!=\!1, \,\, \delta n(\mathbf{R})\!\ll\!1, \,\, \ensavg{\delta n(\mathbf{R})}=0.
\end{equation}
where $\mathbf{R}=[x,y,z]^{T}$ is the three-dimensional position vector, $n(\mathbf{R})$ denotes the the refractive index at $\mathbf{R}$, and $\delta n(\mathbf{R})\!=\!n(\mathbf{R}) - \langle n(\mathbf{R})\rangle$ denotes the small refractive index fluctuation (i.e., optical turbulence)~\citeApndx{book_Apndx}.

The strength of the optical turbulence within a satellite-based atmospheric channel can be described by the structure parameter $C_n^2(h)$ as a function of altitude $h$. $C_n^2(h)$ can be described by the widely used Hufnagel-Valley (HV) model~\citeApndx{book_Apndx}
\begin{equation}\label{Eq.HV}
	\begin{aligned}
		C_{n}^{2}(h)& = 0.00594(v_{\text{rms}}/27)^{2}(h \times 10^{-5})^{10} \exp{(-h/1000)}\\
		&+2.7\!\times\!10^{-16} \exp{(-h/1500)}\!+\!A \exp{(-h/100)},
	\end{aligned}
\end{equation}
where $A$ is the ground-level (i.e., sea-level, $h=0$) turbulence strength in $\text{m}^{-2 / 3}$, and $v_{\text{rms}}$ is the root-mean-square {{(rms)}} wind speed in m/s. We adopt the modified von Karman model for the phase Power Spectral Density (PSD) function of the atmospheric turbulence (see Ref.~\cite{Ziqing_OAMQKD} of the main text and the references therein for more details).

Under the paraxial approximation, the propagation of a monochromatic transverse optical field $\psi(\mathbf{R})$ through the turbulent atmosphere is governed by the stochastic parabolic equation~\citeApndx{book_Apndx}
\begin{equation}\label{sHe}
	\nabla_{\text{T}}^{2} \psi(\mathbf{R})+i 2 k_0 \frac{\partial\psi(\mathbf{R})}{\partial {z}} +2 \delta n(\mathbf{R}) k_0^{2} \psi(\mathbf{R})=0,
\end{equation}
where $k_0=\omega_0/c$ with $c$ being the speed of light in vacuum, and $\nabla_{\text{T}}^{2}=\partial^{2} / \partial x^{2}+\partial^{2} / \partial y^{2}$ is the transverse Laplacian operator.
We adopt the \textit{split-step method}~\citeApndx{SimAP_Apndx} (also  known as the phase screen simulation) to numerically solve the stochastic parabolic equation. This allows us to capture the effects of optical turbulence within the satellite-to-Earth channel on the evolution of the transverse optical field (see Ref.~\cite{Ziqing_OAMQKD} of the main text, \citeApndx{EduardoEnhanced2021_Apdnx}, and the references therein for details).

Note that, the simulation routings for transverse field evolution in the spatial domain are set exactly the same as Ref.~\cite{Ziqing_OAMQKD} of the main text in order to faithfully capture the effects of optical turbulence within a typical satellite-to-Earth atmospheric channel.


\section{ Quantum State Sorting}
Here, we discuss the practical feasibility of quantum state sorting in a satellite-based quantum communication system. Specifically, we discuss the limitations on the practical feasibility of OAM state sorting and TM state sorting within the far-field regime, and we show that the system settings adopted in this work largely avoid such limitations.

It has been discussed (in e.g.,~\citeApndx{ShapiroFarField_Apdnx}) that the channel’s power-transfer behavior has far-field and near-field regimes that are characterized by the Fresnel number product $D_{\text{f}}={A_{\text{T}}}{A_{\text{R}}}/(\lambda L)^2$ with ${A_{\text{T}}}$ ($A_{\text{R}}$) being the transmitter (receiver) aperture area.
Specifically, the near-field regime is defined by $D_{\text{f}}\gg 1$ and the far-field regime is defined by $D_{\text{f}}\ll 1$.
In the far-field regime only one spatial mode (the fundamental Gaussian mode in our case) is supported by the channel (i.e., has significant transmissivity). Intuitively, this  makes sorting OAM states practically impossible even with a perfect OAM sorter. The far-field power-transfer behavior also limits the practical feasibility of TM sorting. Although a TM-based system does not require the use of multiple spatial modes, the received signal power approaches zero as the channel distance $L$ approaches infinity. In this scenario the received signal is dominated by background noise, making TM state sorting impossible regardless of how good the TM sorter is.

The system settings adopted in this work (a LEO satellite and a reasonably large receiver aperture at the ground station) are exactly the same as those used in Ref.~\cite{Ziqing_OAMQKD} of the main text. Our brief estimation indicates that these system settings give $D_{\text{f}} \!\sim\!0.7$ ($\sim\!\!12$) for $r_a=1\,\text{m}$ ($4\,\text{m}$). This means that the satellite-to-Earth quantum communication system considered in this work works within a regime between the near-field regime and the far-field regime.
Therefore, the practical feasibility of both OAM state sorting and TM state sorting is not heavily limited by the far-field power-transfer behavior.

As a satellite-based quantum communication system moves further into the far-field regime (this can be due to the use of a smaller transmitter/receiver aperture or/and the use of a higher satellite altitude),  quantum state sorting becomes more difficult due to the limitations imposed by the far-field power-transfer behavior.
Specifically, the practical feasibility of quantum state sorting in a system working strictly within the far-field regime (e.g., a system using Geosynchronous-Equatorial-Orbit satellites) will be  limited.

\bibliographystyleApndx{IEEEtran}


\end{multicols*}
\end{document}